\documentclass[12pt,mf,harvard]{article}
\usepackage{amssymb}
\usepackage{amsfonts}
\usepackage{amsthm}
\usepackage{stmaryrd}
\usepackage{color}
\usepackage{epsfig}
\usepackage{graphics}
\usepackage[utf8]{inputenc}
\usepackage{hyperref}

\usepackage[
backend=biber,
style=apa,
sorting=nyt
]{biblatex}
\DeclareSourcemap{
  \maps[datatype=bibtex]{
    \map{
      \step[fieldset=issn, null]
      \step[fieldset=doi, null]
      \step[fieldset=url, null]
      \step[fieldset=urldate, null]
    }
  }
}

\hypersetup{
    colorlinks=true,
    urlcolor=blue,
    linkcolor=blue,
    citecolor=blue
}

\newtheorem{theorem}{Theorem}
\newtheorem{lemma}{Lemma}
\newtheorem{proposition}{Proposition}
\newtheorem{assumption}{Assumption}

\newtheorem{example}{Example}
\newtheorem{pilot}{Pilot Example}
\newenvironment{continued}[1][continued]{\begin{trivlist}
\item[\hskip \labelsep {\bfseries #1}]}{\end{trivlist}}
\newtheorem{definition}{Definition}
\newtheorem{remark}{Remark}
\newenvironment{varproof}[1][Proof]{\begin{trivlist}
\item[\hskip \labelsep {\bfseries #1}]}{\end{trivlist}}

\begin{document}
\begin{center}
{\Large\bf A Test of  Non-Identifying Restrictions and Confidence
Regions for Partially Identified Parameters} \vskip30pt {\large
Alfred Galichon and Marc Henry}

\vspace{8pt}
\'Ecole polytechnique, Paris and Universit\'e de Montr\'eal
 \vskip5pt First draft:
September 15, 2005\\This draft\footnotemark[1]:  April 16, 2008
\end{center}

\vskip20pt \footnotetext[1]{This research was partly carried out
while the first author was visiting the Bendheim Center for Finance,
Princeton University and financial support from NSF grant SES
0350770 to Princeton University, from NSF grant SES 0532398, from
the Program for Economic Research at Columbia University and from
Chaire EDF-Calyon ``Finance et D\'eveloppement Durable'' is gratefully acknowledged. We
are grateful to Victor Chernozhukov, Pierre-Andr\'e
Chiappori, Guido Imbens and Bernard Salani\'e for encouragement,
support and many helpful discussions. We also thank three anonymous referees,
whose detailed and insightful comments
helped significantly improve the paper, and we thank
conference participants
at Econometrics in Rio and seminar participants at Berkeley,
Chicago, Columbia, \'Ecole polytechnique, Harvard-MIT, MIT Sloane
OR, Northwestern, NYU, Princeton, SAMSI, Stanford, the Weierstrass
Institut and Yale for helpful comments (with the usual disclaimer).
Correspondence address: D\'epartement d'\'economie, \'Ecole polytechnique, 91128
Palaiseau, France and D\'epartement de sciences \'economiques, Universit\'e de Montr\'eal,
C.P. 6128, succursale Centre-ville, Montr\'eal QC H3C 3J7, Canada.
E-mail: alfred.galichon@polytechnique.edu and marc.henry@umontreal.ca.}

\begin{abstract}
We propose an easily implementable test of the validity of a set of theoretical restrictions on the relationship between economic variables, which do not necessarily identify the data generating process. The restrictions can be derived from any model of interactions, allowing censoring and multiple equilibria. When the restrictions are parameterized, the test can be inverted to yield confidence regions for partially identified parameters, thereby complementing other proposals, primarily \cite{CHT:2007}.
\end{abstract}

\vskip6pt \noindent {\scriptsize JEL Classification: C10, C12,
C13, C14, C52, C61
}

\section*{Introduction}

In several rapidly expanding areas of economic research, the
identification problem is steadily becoming more acute. In policy
and program evaluation (\cite{Manski:90}) and more general
contexts with censored or missing data (\cite{Molinari:2003},
\cite{MM:2005}) and measurement error (\cite{CHT:2005}),
ad hoc imputation rules lead to fragile inference. In demand
estimation based on revealed preference (\cite{BBC:2005}) the
data is generically insufficient for identification. In the analysis
of social interactions (\cite{BD:2005},
\cite{Manski:2004}), complex strategies to reduce the large
dimensionality of the correlation structure are needed. In the
estimation of models with complex strategic interactions and
multiple equilibria (\cite{Tamer:2003}, \cite{ABJ:2003},
\cite{PPHI:2004}), assumptions on equilibrium selection
mechanisms may not be available or acceptable.

More generally, in all areas of investigation with structural data
insufficiencies or incompletely specified economic mechanisms, the
hypothesized structure fails to identify a unique possible
generating mechanism for the data that is actually observed. Hence,
when the structure depends on unknown parameters, and even if a
unique value of the parameter can still be construed as the true
value in some well defined way, it does not correspond in a
one-to-one mapping with a probability measure for the observed
variables. We then call the structural restrictions non-identifying.
In other words, even if we abstract from sampling uncertainty and
assume the distribution of the observable variables is perfectly
known, no unique parameter but a whole set of parameter values
(hereafter called identified set in the terminology of
\cite{Manski:2005}) will be compatible with it.

Once a theoretical description of an economic system is given, a
natural question to consider is whether the structure can be
rejected on the basis of data on its observable components.
\cite{MA:44} construct a collection of production functions
that are compatible with structural restrictions and are not
rejected by the data. We extend this approach within the general
formulation of \cite{KR:50}, who define a structure as the
combination of a binary relation between observed socioeconomic
variables (market entry, insurance coverage, winning bids in
auctions, etc...) and unobserved ones (productivity shocks, risk
level, or risk attitude, valuations or information depending on the
auction paradigm, etc...) and a generating mechanism for the
unobserved variables. This setup is employed by
\cite{Roehrig:88} and \cite{Matzkin:94}, who analyze
conditions for nonparametric identification of structures where the
endogenous observable variables are functions of unobservable
variables and exogenous observable ones.

Here, following \cite{Jovanovic:89}, we allow the relation
between observable and unobservable variables to be many-to-many,
thereby including structures with multiple equilibria (when a value
of the latent variables is associated with a set of values of the
observable variables) and censored endogenous observable variables
(where a value of the observable variable is associated with  set of
values of the latent variables). We do not strive for identification
conditions, but rather for the ability to reject such structures
that are incompatible with data, as in the original work of
\cite{MA:44}.

We show that such a goal can be attained in all generality (ie. for
any structure, involving discrete as well as continuous observable
variables), through an appeal to the duality of mass transportation
(see \cite{Villani:2003} for a comprehensive account of the
theory). Given any set of (possibly non-identifying) restrictions on
the relation between latent and observable variables, and given the
distribution $\nu$ of latent variables, the structure thus defined
is compatible with the true distribution $P$ of the observable
variables if and only if there exists a joint distribution with
marginals $P$ and $\nu$ and such that the restrictions are almost
surely respected. Otherwise, the data could not have been generated
in a such a way. We show that the latter condition can be formulated
as a mass transportation problem (the problem of transporting a
given distribution of mass from an initial location to a different
distribution of mass in a final location while minimizing a certain
cost of transportation, as originally formulated by
\cite{Monge:1781}). We show that this optimization problem has
a dual formulation, an empirical version of which is a generalized
Kolmogorov-Smirnov test statistic. We base a test of the
restrictions in the structure on this statistic, whose asymptotic
distribution we derive, and approximate using the bootstrapped
empirical process.

Once we have a test of the structure, we can form confidence regions
for unknown parameters using the methodology of \cite{AR:49},
which consists in collecting all parameter values for which the
structure is not rejected by the test at the desired significance
level. The construction of such confidence regions has been the
focus of much research lately (see for instance the thorough
literature review in \cite{CHT:2007}). Unlike much of the
econometric research on this issue, we do not restrict the analysis
to models defined by moment inequalities. On the other hand, we
consider structures in the sense of \cite{KR:50}, and hence
parametric distributions for the latent variables. This, however, is
a common assumption in empirical work with game theoretic models, as
exemplified by \cite{ABJ:2003}, \cite{CT:2006}, and more
generally \cite{ABBP:2007}.

The paper is organized as follows. The next section is divided in
four subsections. The first describes the setup; the second defines
the hypothesis of compatibility of the structure with the data; the
third explains how to construct a confidence region for the
identified set, and the fourth reviews the related literature. The
second section is divided in three subsections. The first subsection
describes and justifies the generalized Kolmogorov-Smirnov test of
compatibility of the structure with the data; the second shows
consistency of the test, and the third investigates size properties
of the test in a Monte Carlo experiment. The last section concludes.

\section{Incomplete model specifications}

\subsection{Description of the framework}
Consider the model of an economy which is composed of an observed
variable $Y$ and a latent, unobserved variable $U$. Formally,
$(Y,U)$ is a pair of random vectors defined on a common probability
space. The pair $(Y,U)$ has probability law $\pi$ which is unknown.
$Y$ represents the variables that are observable, and $U$ the
variables that are unobservable. $Y$ may have discrete and
continuous components. $Y$ may include variables of interest in
their own right, and randomly censored or otherwise transformed
versions of variables of interest. We call the law of the observable
variables $P$. It is unknown, but the data available is a sample of
independent and identically distributed vectors $(Y_1,\ldots,Y_n)$
with law $P$. $U$ includes random shocks and other unobserved
heterogeneity components. The law $\pi$ of $(Y,U)$ can be decomposed
into the unconditional distribution $P$ of $Y$ and the conditional
distribution of $U$ given $Y$, namely $\pi_{U\vert Y}$. Throughout
the paper it is supposed that $\pi_{U\vert Y}$  is unknown but fixed
across observations.

The distribution of $U$ is parameterized by a vector
$\theta_1\in\Theta_1$, where $\Theta_1$ is an open subset of
$\mathbb{R}^{d_1}$, and the law of $U$ is denoted $\nu_{\theta_1}$.
Finally, an economic model is given to us in the form of a set of
restrictions on the vector $(Y,U)$, which can be summarized without
loss of generality by the relation $U\in\Gamma_{\theta_2}(Y)$ where
$\Gamma_{\theta_2}$ is a many-to-many mapping, which is completely
given except for the vector of structural parameters
$\theta_2\in\Theta_2$, where $\Theta_2$ is an open subset of
$\mathbb{R}^{d_2}$. $\theta_1$ and $\theta_2$ may contain common
components. We call $\theta$ the combination of the two, so that
$\theta\in\Theta$, with $\Theta$ an open subset of
$\mathbb{R}^{d_\theta}$, and $d_{\theta}\leq d_1+d_2$. From now on,
we shall therefore denote the distribution of $U$ by $\nu_\theta$
and the many-to-many mapping by $\Gamma_\theta$. In all that
follows, we assume that $\Gamma_\theta$ is measurable (a very weak
requirement which is defined in the appendix), and has non-empty and
closed values.

We are interested in testing the compatibility of the observed
variables $Y$ with the model described by $(\Gamma,\nu)$. A  related
question is set-inference in a parametric model
$(\Gamma_\theta,\nu_\theta)$: a confidence region for  $\theta$ can
be obtained by inverting the specification test, namely retaining
the values of $\theta$ which are  not rejected. Note that if
$\theta_2=(\beta,\eta)$, where $\beta$ are the parameters of
interest and $\eta\in H$ are nuisance parameters, we can redefine
the economic model restrictions as $U\in \Gamma_\beta(Y)$ where
$\Gamma_\beta$ is defined by $\Gamma_\beta(y)=\bigcup_{\eta\in
H}\Gamma_{(\beta,\eta)}(y)$ for all $y\in\mathbb{R}^{d_y}$. Hence we
can assume again without loss of generality that $\theta_2$ is
indeed the parameter of interest. As the main focus of the present
paper is to derive a specification test, whenever there is no
ambiguity we shall implicitly fix the parameter $\theta$ and drop it
from our notations.

\begin{example} \label{example: game}
A prominent example for this set-up is provided by the class of
models defined by a static game of interaction. Consider a game
where the payoff function for player $j$, $j=1,\ldots,J$ is given
by $\Pi_j(S_j,S_{-j},X_j,U_j;\theta)$, where
$S_j$ is player $j$'s strategy and $S_{-j}$ is their opponents'
strategies. $X_j$ is a vector of observable characteristics of
player $j$ and $U_j$ a vector of unobservable determinants of the
payoff. Finally $\theta$ is a vector of parameters. Pure strategy
equilibrium conditions
define a many-to-many mapping $\Gamma_{\theta}$ from unobservable
player characteristics $U$ to observable variables $Y=(S,X)$. More
precisely,
$\Gamma_\theta(s,x)=\{u\in\mathbb{R}^{J}:\;\Pi_j(s_j,s_{-j},x_j,u_j;\theta)\geq
\Pi_j(s,s_{-j},x_j,u_j;\theta),\mbox{ for all }S\mbox{ and all
}j\}$. When the strategies are discrete, this is the set-up
considered by \cite{ABJ:2003}, \cite{PPHI:2004}, and
\cite{CT:2006}.
\end{example}

A special case of the latter example is given in
\cite{Jovanovic:89} and will serve as our first illustrative example:

\begin{pilot}\label{pilot: cooperative game}
The payoff functions are
$\Pi_1(Y_1,Y_2,U_1,U_2)=(\theta Y_2-U_2)1_{\{Y_1=1\}}$ and
$\Pi_2(Y_1,Y_2,U_1,U_2)=(\theta Y_1-U_1)1_{\{Y_2=1\}}$,
where $Y_i\in\{0,1\}$ is firm i's action, and $U=(U_1,U_2)'$ are
exogenous costs. The firms know their costs; the analyst, however,
knows only that $U$ is uniformly distributed on $[0,1]^2$, and that
the structural parameter $\theta$ is in $(0,1]$. There are two pure
strategy Nash equilibria. The first is $Y_1=Y_2=0$ for all
$U\in[0,1]^2$. The second is $Y_1=Y_2=1$ for all $U\in[0,\theta]^2$
and zero otherwise. Since the two firms' actions are perfectly
correlated, we shall denote them by a single binary variable
$Y=Y_1=Y_2$. Hence the structure is described by the many-to-many
mapping: $\Gamma_\theta(1)=[0,\theta]$ and $\Gamma_\theta(0)=[0,1]$.
In this case, since $Y$ is Bernoulli, we can characterize $P$ with
the probability $p$ of observing a 1.\end{pilot}

A second example illustrates the case with continuous observable variables:

\begin{pilot}\label{pilot: labour market}
\cite{Tinbergen:51} first spelt out the implications of skill and job requirement heterogeneity on the distribution of wages.
We adopt a simplified version of the skill versus job requirements relation for illustrative purposes.
Suppose one observes available jobs in an economy, each characterized
by a set of characteristics $Y$ with distribution $P$. Worker's skills are unobserved, and are
assumed for illustrative purposes to be characterized by an index $U\in\mathbb{R}$.
Fulfillment of job $Y$ is known to require a range of skills $\Gamma_\theta(Y)=
[\underline{s}_\theta(Y),\overline{s}_\theta(Y)]$. The distribution of skills
is parameterized by $\nu_\theta$. \end{pilot}

\subsection{Partial Identification}

Identification of the parameter $\theta$ would require the
correspondence between the law of the observations $P$ and the
parameter vector $\theta$ to be a function. Compared to the setup
described in \cite{Roehrig:88}, there is the added complexity
of the possibility that the observable variables have discrete
components, and that the structure allows multiple equilibria.
Conditions ensuring identification are likely to prove complicated
and restrictive, and will often rule out multiple equilibria, which
is the norm rather than the exception in example~\ref{example:
game}. We therefore eschew identification, and allow the relation
between $P$ and $\theta$ to be many-to-many. Our objective is to
conduct inference on the set $\Theta_I$ of parameter values that are
compatible with the true law of the observable variables $P$.

Let us formally define compatibility of a given value $\theta_0$ of
the parameter vector with a law $P$ for the observable variables
$Y$. When $\theta_0$ is fixed, all the elements in the model are
completely known. We therefore have a structure in the terminology
of \cite{KR:50} extended by \cite{Jovanovic:89}. The
structure is given by the law $\nu_{\theta_0}$ for $U$, and the
many-to-many mapping $\Gamma_{\theta_0}$ linking $Y$ and $U$. We
denote this structure by the triple
$(P,\Gamma_{\theta_0},\nu_{\theta_0})$. Consider now the
restrictions that $(P,\Gamma_{\theta_0},\nu_{\theta_0})$ imposes on
the unknown $\pi$, the law of the vector of variables $(Y,U)$.
\begin{itemize} \item Its marginal with respect to $Y$ is $P$,
\item Its marginal with respect to $U$ is $\nu_{\theta_0}$, \item The
economic restrictions $U\in\Gamma_{\theta_0}(Y)$ hold $\pi$ almost
surely.
\end{itemize}
A probability law $\pi$ that satisfies the restrictions above may or
may not exist. If and only if it does, we say that the structure
$(P,\Gamma_{\theta_0},\nu_{\theta_0})$ is internally consistent, or
simply that the value $\theta_0$ of the parameter is compatible with
the law $P$ of the observable variables. If no value $\theta_0$ is
found such that the structure is internally consistent, then the
model restrictions are rejected.

\begin{definition}\label{definition: internal consistency}
A structure $(P,\Gamma,\nu)$ for $(Y,U)$ given by a probability law $P$
for $Y$, a probability law $\nu$ for $U$ and a set of restrictions
$U\in\Gamma(Y)$ is called internally consistent if there exists a
law $\pi$ for the vector $(Y,U)$ with marginals $P$ and $\nu$ such that
$\pi(\{U\in\Gamma(Y)\})=1$. \end{definition}

We can now define the identified set as the set of values of the
parameters that achieve this internal consistency. They are
observationally equivalent, since even though they may correspond
to different $\pi$'s, they correspond to the same $P$.

\begin{definition}\label{definition: identified set}
The identified set $\Theta_I=\Theta_I(P)$ is the set of values
$\theta$ of the parameter vector such that the structure
$(P,\Gamma_\theta,\nu_\theta)$ is internally consistent.
\end{definition}

We illustrate the previous definitions with our pilot example:

\begin{continued}[Pilot example \protect\ref{pilot: cooperative game} continued]
For a given value of $\theta$, the structure
$(P,\Gamma_\theta,\nu_\theta)$ is defined by $p$, $\Gamma_\theta$
and the uniform distribution $\nu_\theta$ on $[0,1]^2$.
$(P,\Gamma_\theta,\nu_\theta)$ is internally consistent if there
exists a probability on $\{0,1\}\times[0,1]^2$ with marginal
frequency $p$ of observing a $Y=1$, and uniform marginal
distribution for the costs $U$ such that $Y=1\Rightarrow
U\leq\theta$ almost surely (where the last inequality is meant
coordinate by coordinate).
\end{continued}

The previous example illustrates the fact that definition~\ref{definition: internal consistency}
is not very easy to apply to derive the identified set in specific problems. We therefore propose a
characterization of internal consistency which will prove more practical, and which, as we shall see in the next section,
will motivate the construction of the statistic to test internal consistency.

\begin{proposition}
\label{proposition: Monge-Kantorovich} A structure $(P,\Gamma,\nu)$
is internally consistent if and only if $\sup_{A\in{\cal B}}
[P(A)-\nu(\Gamma(A))]=0$ where ${\cal B}$ is the collection of
measurable sets in the space of realizations of $Y$.
\end{proposition}

This proposition shows that checking internal consistency of a
structure is equivalent to checking that the $P$-measure of a set is
always dominated by the $\nu$-measure of the image of this set by
$\Gamma$ (recall that the image of a set $A$ by a many-to-many
mapping is defined by $\Gamma(A)=\bigcup_{a\in A}\Gamma(a)$). Note
that it is relatively easy to show necessity, i.e. that the
existence of $\pi$ satisfying the constraints (the definition of
internal consistency) implies that $\sup_{A\in{\cal B}}
[P(A)-\nu(\Gamma(A))]=0$. Indeed, the definition of internal
consistency implies that ${Y\in A}\Rightarrow U\in\Gamma(A)$, so
that $1_{\{Y\in A\}}\leq1_{\{U\in\Gamma(A)\}}$, $\pi$-almost surely.
Taking expectation, we have $\mathbb{E}_\pi(1_{\{Y\in A\}})\leq
\mathbb{E}_\pi(1_{\{U\in\Gamma(A)\}})$, which yields the result,
since $\pi$ has marginals $P$ and $\nu$. The converse (proved in the
appendix) is far more involved, as it relies on mass transportation
duality, where mass $P$ is transported into mass $\nu$ with 0-1 cost
of transportation associated with violations of the restrictions
$U\in\Gamma(Y)$.

\begin{continued}[Pilot example \protect\ref{pilot: cooperative game} continued]
For a given $\theta$, it is now very easy to derive the condition
for internal consistency of the structure. Indeed, all we need to
check is that $\sup_{A\in2^{\{0,1\}}}[P(A)-
\nu_\theta(\Gamma_\theta(A))]=0$ (where $2^B$ is the collection of
all subsets of a set $B$), which only constrains
$P(\{1\})\leq\nu_\theta([0,\theta]^{2})$, hence $p\leq\theta^2$. So
the identified set for the structural parameter is
$\Theta_I=[\sqrt{p},1]$.
\end{continued}

\begin{remark} Further dimension reduction requires the determination of
classes of sets $A$ on which to check the inequality between $P(A)$
and $\nu(\Gamma(A))$. This is needed for instance when the
observable variables are discrete and take many different values,
since checking the inequality for all subsets of the set of possible
values would involve a very large number of operations.
\cite{GH:2006a} addresses this issue with a theory of {\em
core determining classes}.\end{remark}

\begin{continued}[Pilot example \protect\ref{pilot: labour market} continued]
Fixing $\theta$ (and dropping it from the notation), the necessary
and sufficient condition for internal consistency of the structure
is that $P(A)\leq\nu(\Gamma(A))$ for any measurable set $A$. Suppose
for expositional purposes that the jobs  are characterized by a real
valued random variable $Y$, and that required skills are monotone in
the sense that $\underline{s}$ and $\overline{s}$ are nondecreasing.
As shown in \cite{GH:2006a},
the inequality needs to be checked only on sets of the form
$A=(-\infty,y]$ and $A=(y,+\infty)$, for $y\in\mathbb{R}$, so that a
necessary and sufficient condition for internal consistency of the
structure is that $F_\nu(\underline{s}(y))\leq F(y)\leq
F_\nu(\overline{s}(y))$, where $F$ is the cumulative distribution
function of jobs $Y$, and $F_\nu$ is the cumulative distribution
function of skills $U$.
\end{continued}

\subsection{Inference on the identified set}
Given a sample $(Y_1,\ldots,Y_n)$ of independently and identically
distributed realizations of $Y$, our objective is to construct a
sequence of random sets $\Theta_{n}^{\alpha}$ such that for all
$\theta\in\Theta_I$,
$\lim_{n\rightarrow\infty}\mbox{Pr}\left(\theta\in\Theta_n^{\alpha}\right)=
1-\alpha$. In other words, we are concerned with constructing a
region $\Theta_n^{\alpha}$ that covers each value of the identified
set, as opposed to a region $\tilde\Theta$ that covers the
identified set uniformly, i.e. such that
Pr$(\Theta_I\subseteq\tilde{\Theta})=1-\alpha$. We do so by
including in $\Theta_n^{\alpha}$ all the values of $\theta$ such
that we fail to reject a test of internal consistency of
$(P,\Gamma_\theta,\nu_\theta)$ with asymptotic level $1-\alpha$. We
shall demonstrate the construction of a test statistic $T_n(\theta)$
and a sequence $c_n^\alpha(\theta)$ such that, conditionally on the
structure $(P,\Gamma_\theta,\nu_\theta)$ being internally
consistent, the probability that $T_n(\theta)\leq
c_n^\alpha(\theta)$ is  $1-\alpha$ asymptotically, i.e.
\begin{eqnarray}\lim_{n\rightarrow\infty} \mbox{Pr}\left(
T_n(\theta)\leq c_n^\alpha(\theta)\;\vert\;(P,\Gamma_\theta,\nu_\theta)\; \mbox{is
internally consistent} \right)=1-\alpha.\label{equation:
size}\end{eqnarray} Hence we define our confidence region in the
following way.

\begin{definition}The $(1-\alpha)$ confidence region for $\Theta_I$ is $\Theta_n^{\alpha}=\{
\theta\in\Theta: T_n(\theta)\leq c_n^\alpha(\theta)\}$.\end{definition}

The full procedure is summarized in table~1. It is clear from
equation~\ref{equation: size} and the above definition that our
confidence region covers each element of the identified set with
probability $1-\alpha$ asymptotically. Hence, after a section
devoted to discussing in detail our contribution within the
literature on the topic, the remainder of this paper will be
concerned with the construction of the statistic $T_n$ and sequence
$c_n^\alpha$ with the required property~(\ref{equation: size}).

\begin{table} \label{table: summary}
\begin{center}
\caption{\bf Summary of the procedure} \vskip15pt \frame{
\begin{minipage}[b]{5.1in}
\vskip5pt \small\em \begin{itemize}  \item[1.] For a given value of
$\theta$, calculate $\hat{T}_n(\theta)=\sqrt{n}\sup_{A\in{\cal C}_n}
[P_n(A)-\nu_\theta(\Gamma_\theta(A))]$, where the collection of sets
${\cal C}_n$ is described in table~2, and $P_n$ is the empirical
distribution of the sample $(Y_1,\ldots,Y_n)$, so that
$P_n(A)=(1/n)\sum_{i=1}^n1_{\{Y_i\in A\}}$.
\item[2.] Choose a large integer $B$.
Draw $B$ bootstrap samples $(Y_1^b,\ldots,Y_n^b)$, $b=1,\ldots, B$
with replacement from the initial sample $(Y_1,\ldots,Y_n)$. For
each bootstrap sample, calculate $T^b_n(\theta)=\sup_{A\in{\cal
C}_{n,h_n}(\theta)}[P^b(A)-P_n(A)]$, where $P^b$ is the empirical
distribution of the bootstrap sample, and ${\cal C}_{n,h_n}(\theta)$
is described in table~2. Order the $T^b_n(\theta)$'s and call
$c_\ast^\alpha(\theta)$ the $B(1-\alpha)$ largest. \item[3.] Include
$\theta$ in $\Theta_I$ if and only if $\hat{T}_n(\theta)\leq
c_\ast^\alpha(\theta)$.
 \end{itemize}  \vskip5pt
\end{minipage}
}
\end{center}
\end{table}

\begin{table} \label{table: collections of sets}
\begin{center}
\caption{\bf Collection of sets} \vskip15pt \frame{
\begin{minipage}[b]{5.1in}
\vskip5pt \small\em \begin{itemize}  \item[1.] Take the sample
$(Y_1,\ldots,Y_n)$. Write $Y_i=(D_i,C_i)$ where $D_i$ includes the
discrete components, and $C_i$ the continuous components of the
observable variables in the sample. Call $\mathfrak{X}_D$ the set of
values taken by $D_i$. Then, ${\cal C}_n$ is the collection of sets
of the form $A_D\times[-\infty,C_i]$ or its complement, where
$i=1,\ldots,n$, $A_D$ ranges over the subsets of $\mathfrak{X}_D$,
and $[-\infty,C_i]$ denotes the hyper-rectangle bounded above by the
components of~$C_i$. \item[2.] Given $h_n$ satisfying $h_n\ln\ln n +
h_n^{-1}\sqrt{\ln\ln n/n}\rightarrow0$ as $n\rightarrow\infty$ (e.g.
$h_n=(\ln n)^{-1}$), take
$\mathcal{C}_{n,h_n}(\theta)=\{A\in\mathcal{C}_n:\;
P_n(A)\geq\nu_\theta(\Gamma_\theta(A))-h_n\}$.
 \end{itemize}  \vskip5pt
\end{minipage}
}
\end{center}
\end{table}

\begin{continued}[Pilot example \protect\ref{pilot: cooperative game} continued]
The test statistic is then
$T_n(\theta)=\sqrt{n}\sup_{A\in2^{\{0,1\}}}
[P_n(A)-\nu_\theta(\Gamma_\theta(A))]$. Since
$P_n(\varnothing)=\nu_\theta(\Gamma_\theta(\varnothing))$ and
$P_n(\{1\})-\nu_\theta(\Gamma_\theta(\{1\}))=p_n-\theta^2$, the test
statistic is equal to
$T_n(\theta)=\max\{\sqrt{n}(p_n-p)+\sqrt{n}(p-\theta^2),0\}$ which
tends to $\max\{\sqrt{p(1-p)}Z,0\}$ where $Z$ is a standard normal
random variable, if $p=\theta^2$, $0$ if $p<\theta^2$, $+\infty$ if
$p>\theta^2$. For any $\theta$ such that $p\leq\theta^2$,
$T_n(\theta)$ has the same limit as
$\tilde{T}_n=\sup_{A\in\mathcal{C}_{h_n}}[\sqrt{n}(P_n(A)-P(A))]$
where $\mathcal{C}_{h_n}$ is equal to $\{\varnothing,\{0,1\}\}$ if
$p_n<\theta^2-h_n$ and $2^{\{0,1\}}$ if $p_n\geq\theta^2-h_n$. Hence
the confidence region $\Theta^\alpha_n$ is the set of $\theta$
values that are not rejected in a one-sided test of the null
hypothesis $p\leq\theta^2$ against the alternative $p>\theta^2$
based on the quantiles of the distribution of
$\max\{\sqrt{n}(p^\ast-p_n),0\}$ given the sample (where $p^\ast$
denotes the frequency of $1$'s in a bootstrap sample).

\end{continued}

\subsection{Review of the literature}
This paper appears to be the first to cast partial identification as
a mass transportation problem. Somewhat related is the specific use
of Fr\'echet-Hoeffding bounds on cell probabilities in
\cite{HSC:97} and \cite{CM:2002}.

The literature on specification testing in econometrics is quite
extensive (see the many references in \cite{Andrews:88} for
Cram\'er-von Mises tests and \cite{Andrews:97} for the
Kolmogorov-Smirnov type). \cite{Jovanovic:89} proposes to
consider testing specifications with multiple equilibria and
possible lack of identification with a generalization of the
Kolmogorov-Smirnov specification test, which is exceedingly
conservative unless the structure is nearly identified. The
stochastic dominance tests of \cite{McFadden:89} (see also
\cite{LMW:2005} and references within) are also related to
tests of partially identified structures based on the
Kolmogorov-Smirnov statistic. The feasible version of our testing
procedure and the use of the bootstrapped empirical process is
related to \cite{Andrews:97}.

The incompleteness of the structure to be tested raises boundary
problems, which appear also in the estimation of models defined by
moment inequalities (see \cite{IM:2004} and the link drawn by
\cite{Rosen:2006} with the literature on constrained
statistical testing, surveyed in \cite{SS:2004}) and
stochastic dominance testing (see \cite{LMW:2005}). Here the
asymptotic analysis is carried out via a localization of the
empirical processes to treat the boundary problem, which is another
major innovation of this paper. Also related is the analysis in
\cite{LS:2003} of the likelihood ratio test when the
likelihood is maximized on a set as opposed to a single point.

The related problem of constructing confidence regions for partially
identified structural parameters is the focus of considerable recent
research, following the recognition (advocated in
\cite{Manski:2005}) that ad-hoc identification conditions can
considerably weaken inference drawn on their basis.
\cite{HM:98} propose confidence intervals that asymptotically
cover interval identified sets with fixed probability. Beyond the
interval case, \cite{CHT:2007} propose a criterion function
based method, where the criterion is maximized on a set, as opposed
to a single point. The method allows the construction of confidence
regions for the identified set and for each parameter value in the
identified set. \cite{CHT:2007} also specialize their method
to the case of models defined by moment inequalities, with a
quadratic criterion function.

The case of moment inequalities is also considered as a special case
by \cite{GH:2006c}, \cite{RS:2006a} and
\cite{RS:2006b} (see also \cite{Rosen:2006} and
\cite{Bugni:2007}). The present paper complements
\cite{CHT:2007} in that it justifies, via a mass
transportation argument, the use of a generalized Kolmogorov-Smirnov
criterion function in the extended \cite{KR:50} setup
presented here. Note that our proposed use of the bootstrap only
concerns the empirical process, as in \cite{Andrews:2000}, so
that issues of validity related to bootstrapping the test statistic
itself do not arise.

The \cite{AR:49} approach taken here to construct confidence
regions for parameter values within the identified set is also
adopted in \cite{CHT:2007}, \cite{ABJ:2003},
\cite{RS:2006a} among many others. \cite{ABJ:2003} work
in a similar framework to the present paper (they consider
example~\ref{example: game}), but restrict their analysis to
discrete dependent variables, and use a projection method, so that
their inference is likely to be more conservative.

Since confidence regions are asymptotically validated, as emphasized
by \cite{IM:2004}, uniformity of the confidence region for
parameter values is a desirable property for small sample accuracy.
\cite{AG:2006} analyze uniformity of sub-sampling procedures.
\cite{RS:2006a} and \cite{RS:2006b} give high level
conditions for uniformity of sub-sampling procedures in the
criterion-based approach, with specific conditions under which these
results hold in case of regression with interval outcomes. Here, we
propose to invert a test, which is shown to be asymptotically
uniform in level in \cite{GH:2008}.

In related research, \cite{BM:2007} propose a direct analogy
to central limit theorem based confidence regions in best linear
prediction problems. The confidence region they propose for the
identified set, in a problem of best linear prediction with interval
outcomes, is the union of a collection of random sets that contain
the identified set with pre-specified probability. The latter is
obtained from central limit theorems for random sets (see
\cite{Molchanov:2005} for a comprehensive account of the
theory). They propose one-sided and two-sided versions of their test.
The \cite{BM:2007} two-sided procedure does not
suffer from discontinuity at the limit where the identified set is a
singleton. However, by construction, \cite{BM:2007} only
provide confidence regions for the whole set, which are typically
larger than identified regions for each point in the identified set.

\section{Test of internal consistency}

As explained in the previous section, the construction of the
confidence region relies on a test of internal consistency of the
structure $ (P,\Gamma_\theta,\nu_\theta)$ for a fixed $\theta$. We
now explain the construction of our test statistic and decision
rule, for the hypothesis of internal consistency of a structure
$(P,\Gamma,\nu)$ defined by a a probability law $\nu$ for $U$ and a
set of constraints $U\in\Gamma(Y)$. The hypothesis that
$(P,\Gamma,\nu)$ is internally consistent is equivalent to the
existence of a law $\pi$ for $(Y,U)$ with marginals $P$ and $\nu$
and such that the constraints $U\in\Gamma(Y)$ hold $\pi$-almost
surely. By proposition~\ref{proposition: Monge-Kantorovich}, this
null hypothesis is also equivalent to
\[\mathbb{H}_0:\;\sup_{A\in{\cal B}}[P(A)-\nu(\Gamma(A))]=0.\]

\subsection{Test statistic and size of the test of internal
consistency} We propose the following statistic to test the null
described above:
\begin{eqnarray} T_n=\sqrt{n}\sup_{A\in\mathcal{C}}[P_n(A)-\nu(\Gamma(A))],
\label{equation: test statistic}\end{eqnarray}
where $P_n$ is the empirical distribution of the sample (so that for
any measurable set $A$, $P_n(A)=(1/n)\sum_{i=1}^n1_{\{Y_i\in A\}}$)
and where $\mathcal{C}$ is defined in table~3.

\begin{table} \label{table: theoretical collection of sets}
\begin{center}
\caption{\bf  collections of sets} \vskip15pt \frame{
\begin{minipage}[b]{5.1in}
\vskip5pt \small\em \begin{itemize}  \item[1.] Write $Y=(D,C)$ where
$D$ includes the discrete components, and $C$ the continuous
components with dimension $d_C$. Call $\mathfrak{X}_D$ the set of
values taken by $D$. Then, ${\cal C}$ is the collection of sets of
the form $A_D\times[-\infty,c]$ or its complement, where
$c\in\mathbb{R}^{d_C}$, $A_D$ ranges over the subsets of
$\mathfrak{X}_D$, and $[-\infty,c]$ is the hyper-rectangle bounded
above by the components of~$c$. \item[2.] Given $h>0$, define
\begin{itemize}
\item[] $\mathcal{C}_{b}=\{A\in\mathcal{C}:\;
P(A)=\nu(\Gamma(A))\}$. \item[]
$\mathcal{C}_{b,h}=\{A\in\mathcal{C}:\; P(A)\geq\nu(\Gamma(A))-h\}$.
\item[] $\mathcal{C}_{h}=\{A\in\mathcal{C}:\;
P_n(A)\geq\nu(\Gamma(A))-h\}$.
\end{itemize}
 \end{itemize}  \vskip5pt
\end{minipage}
}
\end{center}
\end{table}

This statistic is a generalized Kolmogorov-Smirnov specification
test statistic in the sense that when $\Gamma$ has disjoint images (i.e. $\Gamma^{-1}$ is
a function), $T_n$ is a multivariate
Kolmogorov-Smirnov statistic for the test of the hypothesis that the
structure is correctly specified, i.e. that the probability law
$A\mapsto\nu(\Gamma(A))$ is indeed equal to the true law $P$
generating the observable variables $Y$. In the general case where
$\Gamma$ is a many-to-many mapping, $A\mapsto\nu(\Gamma(A))$ is no
longer a probability measure, since two sets $A$ and $B$ may be
disjoint, and yet their images $\Gamma(A)$ and $\Gamma(B)$ are not,
so that $\nu(\Gamma(A\cup B))$ may be strictly smaller than
$\nu(\Gamma(A))+\nu(\Gamma(B))$. This introduces significant
complications in the asymptotic analysis of the statistic $T_n$ as
explained in the following discussion.

We can write
\begin{eqnarray}T_n=\sqrt{n}\sup_{A\in\mathcal{C}}[P_n(A)-\nu(\Gamma(A)]=
\sup_{A\in\mathcal{C}}\{\mathbb{G}_n(A)+\sqrt{n}[P(A)-\nu(\Gamma(A)]\}
\label{equation: empirical process}\end{eqnarray} where
$\mathbb{G}_n(A):=\sqrt{n}[P_n(A)-P(A)]$ is the empirical process.
In the case of the classical Kolmogorov-Smirnov statistic (i.e. if
$\Gamma^{-1}$ were a function), the term $P(A)-\nu(\Gamma(A))$ would
vanish under the null hypothesis. Here, however, under the null we
only have $P(A)\leq\nu(\Gamma(A))$, so that the term
$\sqrt{n}[P(A)-\nu(\Gamma(A)]$ will also contribute. Indeed, for any
set $A\in\mathcal{C}$ such that $P(A)=\nu(\Gamma(A))$ (i.e.
$A\in\mathcal{C}_b$ as defined in table~3), the only remaining term
in the right-hand-side of equation~(\ref{equation: empirical
process}) is the empirical process. On the other hand, for any set
$A\in\mathcal{C}$ such that $P(A)<\nu(\Gamma(A))$,
$\sqrt{n}[P(A)-\nu(\Gamma(A))]$ will take increasingly large
negative values and eventually dominate the expression inside the
supremum in the right-hand-side of equation~(\ref{equation:
empirical process}) and such a set $A$ will not contribute to the
supremum. We show in the proof of theorem~\ref{theorem: size} that
under a very mild assumption on the structure, the limit will only
involve a supremum over sets in $\mathcal{C}_b$. Since
$\mathcal{C}_b$ depends on $P$, it is unknown, and needs to be
approximated by a data dependent class $\mathcal{C}_{h_n}$ defined
in table~3 (namely $\mathcal{C}_h$ with $h=h_n$).

\begin{definition} The test statistic $T_n$ is given by equation~(\ref{equation: test statistic}),
and $c_n^\alpha$ is the $1-\alpha$ quantile of
$\tilde{T}_n:=\sup_{A\in\mathcal{C}_{h_n}}\mathbb{G}_n(A)$
(with $\mathcal{C}_h$ defined in table~3), i.e.
$c_n^\alpha = \inf\{c: \;\mathbb{P}(\tilde{T}_n\leq c) \geq 1-\alpha\}$.
\label{definition: test statistic}\end{definition}

\begin{assumption} There exists
$K>0$ and $0<\eta<1$ such that for all $A\in{\cal C}_{b,h}$, for
$h>0$ sufficiently small, there exists an $A_b\in{\cal C}_b$ such
that $A_b\subseteq A$ and $d_H(A,A_b)\leq K h^{\eta}$.
($\mathcal{C}_b$ and $\mathcal{C}_{h}$ are defined in table~3, and
$d_H$ denotes the Hausdorff metric, defined in the appendix.)
\label{assumption: frank separation}\end{assumption}

\begin{remark} Assumption~\ref{assumption: frank separation} is very mild, in the sense that it fails
only in pathological cases, such as the case where $y\in\mathbb{R}$
and $y\mapsto P((-\infty,y])-\nu(\Gamma((-\infty,y]))$ is
$C^{\infty}$ with all derivatives equal to zero at some $y=y_0$ such
that $(-\infty,y_0]\in{\cal C}_b$. \end{remark}

\begin{assumption}$h_n$ satisfies $h_n\ln\ln
n+h_n^{-1}\sqrt{\ln\ln n/n}\rightarrow0\;\mbox{ as }
\;n\rightarrow\infty$.\label{assumption: bandwidth}\end{assumption}

\begin{remark} Note that assumption~\ref{assumption: bandwidth} is extremely mild, and it is
satisfied for instance in case $h_n=(\ln n)^{-1}$ or in case $h_n$
satisfies $h_n n^{\eta}+h_n^{-1} n^{\eta-1/2}\rightarrow0$, as
$n\rightarrow\infty$ for any $1/2>\eta>0$, however
small.\end{remark}

\begin{theorem} Suppose $Y$ either
takes values in a finite set or has density with respect to Lebesgue
measure. Under assumption~\ref{assumption: frank separation}
and~\ref{assumption: bandwidth}, and using the notations of
definition~\ref{definition: test statistic}, we have
\[ \lim_{n\rightarrow\infty}\mathbb{P}(T_n\leq c_n^\alpha\;\vert\; (P,\Gamma,\nu)
\mbox{ is internally consistent })=1-\alpha.\]
\label{theorem: size}\end{theorem}

Theorem~\ref{theorem: size} is not applicable directly for
two reasons: \begin{itemize}

\item[1.] The quantile sequence $c_n^\alpha$ given in definition~\ref{definition: test statistic}
is infeasible in that the statistic $\tilde{T}_n$ involves the empirical
process $\mathbb{G}_n=\sqrt{n}[P_n-P]$ with $P$ unknown.

\item[2.] The statistics $T_n$ and $\tilde{T}_n$ are defined as suprema over
infinite collections of sets $\mathcal{C}$ and $\mathcal{C}_h$ (with
$\mathcal{C}$ and $\mathcal{C}_h$ defined in table~3).

\end{itemize}

We show now that $T_n$ can be replaced by $\hat{T}_n$ defined in
table~2, and that $c_n^\alpha$ can be replaced by $c_\ast^\alpha$,
which is the $1-\alpha$ quantile of
$T^\ast:=\sup_{A\in\mathcal{C}_{n,h_n}}\mathbb{G}^\ast(A)$, where
$\mathbb{G}^\ast:=\sqrt{n}[P^\ast-P_n]$ is the bootstrapped
empirical process. We thereby justify the fully implementable
procedure described in table~1. This feasible version of the test
mirrors the feasible version of the conditional Kolmogorov-Smirnov
test proposed by \cite{Andrews:97}, albeit in generalized form
(multivariate and incompletely specified).

To that end, we need a large support assumption and a log concavity
assumption for the distribution of observable variables and a
continuity assumption on the mapping $\Gamma$ to ensure that
$\hat{T}_n$ has the same limit as $T_n$.

\begin{assumption}In case $P$ has density with respect to Lebesgue measure,
the density is bounded away from zero, absolutely continuous and log
concave (note that log concave densities include the uniform,
normal, beta, exponential and extreme value distributions).
\label{assumption: log concave}\end{assumption}

\begin{assumption}
The functions $y\mapsto\nu(\Gamma((-\infty,y]))$ and
$y\mapsto\nu(\Gamma((-\infty,y]^c))$ are Lipschitz, i.e. there
exists some $k>0$ such that
$\vert\nu(\Gamma((-\infty,y]))-\nu(\Gamma((-\infty,y']))\vert\leq
k\vert\vert y-y'\vert\vert$, and identically for $(-\infty,y]^c$.
\label{assumption: continuity}\end{assumption}

\begin{theorem}\label{theorem: feasible procedure}
Under the assumptions of theorem~\ref{theorem: size} and
assumptions~\ref{assumption: log concave} and~\ref{assumption:
continuity}, we have
\[ \lim_{n\rightarrow\infty}\mathbb{P}(\hat{T}_n\leq c_\ast^\alpha\;\vert\; (P,\Gamma,\nu)
\mbox{ is internally consistent })=1-\alpha\] almost surely,
conditionally on the sample.
\end{theorem}

\begin{remark}
The conditions for the validity of the bootstrap procedure are no
more restrictive than the conditions for theorem~\ref{theorem:
size}. The additional assumptions, which are more high level, are
needed only to justify using the data driven class of sets ${\cal
C}_n$ instead of ${\cal C}$. This follows the proposal in
\cite{Andrews:97} in order to simplify the testing procedure
as much as possible. However, an alternative feasible version of the
test relies on a regular discretization $(y_k)_{k=1}^{N}$ of the
space of continuous observable variables (thereby replacing ${\cal
C}_n$ by the class of sets of the form $(-\infty,y_k]$,
$(-\infty,y_k]^c$, $k=1,\ldots,N$).
\end{remark}

\subsection{Consistency of the test}
To complete the analysis of the test of internal consistency we give
conditions under which the test is consistent. The class of
alternatives we consider is the following:
\[\mathbb{H}_a:\;\sup_{A\in{\cal C}}[P(A)-\nu(\Gamma(A))]\neq0,\]
where $\mathcal{C}$ is defined in table~3. We choose this class of
alternatives since it simplifies to the set of alternatives in a
multivariate Kolmogorov-Smirnov goodness-of-fit test when $P$ is
absolutely continuous with respect to Lebesgue measure and when
$\Gamma^{-1}$ is a function.

We have \begin{theorem}Under $\mathbb{H}_a$ and the assumptions of
theorem~\ref{theorem: size},
$\lim_{n\rightarrow\infty}\mathbb{P}(T_n\geq
c_n^\alpha)=1$\label{theorem: consistency}.\end{theorem}

\begin{remark}
Notice that the validity of this consistency test is completely
general, and, unlike theorem~\ref{theorem: size}, the proof is a
straightforward extension of the proof of consistency of the
traditional Kolmogorov-Smirnov specification test (see for instance
page 526 of \cite{LR:2005}).
\end{remark}

\subsection{Small sample investigation of the properties of the test of internal consistency}
We investigate the small sample properties of out test, and compare it to the properties of the Kolmogorov-Smirnov
specification test in the identified case in a small Monte Carlo experiment based on a special case of
illustrative example~\ref{pilot: labour market}.

We consider the following setup illustrated in figure~1: the structure is given by the correspondence $\Gamma(Y)=[\underline{s}(Y),
\overline{s}(Y)]$ with $\underline{s}(Y)=\max(0,Y+s)$ and $\overline{s}(Y)=\min(1,Y+s)$, $s=0.15$,
and the latent variable
$U$ has law $\nu$, which is the uniform distribution over $[0,1]$.
$Y$ has cumulative distribution function defined on $[0,1]$ by
\begin{eqnarray*}F(y)&=&0\;\mbox{ for }\;0\leq y<s,\\
&=&y-s\;\mbox{ for }\;s\leq y<\frac{1+s}{3},\\
&=&\frac{(1+4s)y-3s}{1-2s}\;\mbox{ for }\;\frac{1+s}{3}\leq y<\frac{2-s}{3},\\
&=&y+s\;\mbox{ for }\;\frac{2-s}{3}\leq y<1-s,\\
&=&1\;\mbox{ for }\;1-s\leq y\leq1.
\end{eqnarray*}

\begin{figure}[htbp]
\begin{center}
\includegraphics[width=12cm]{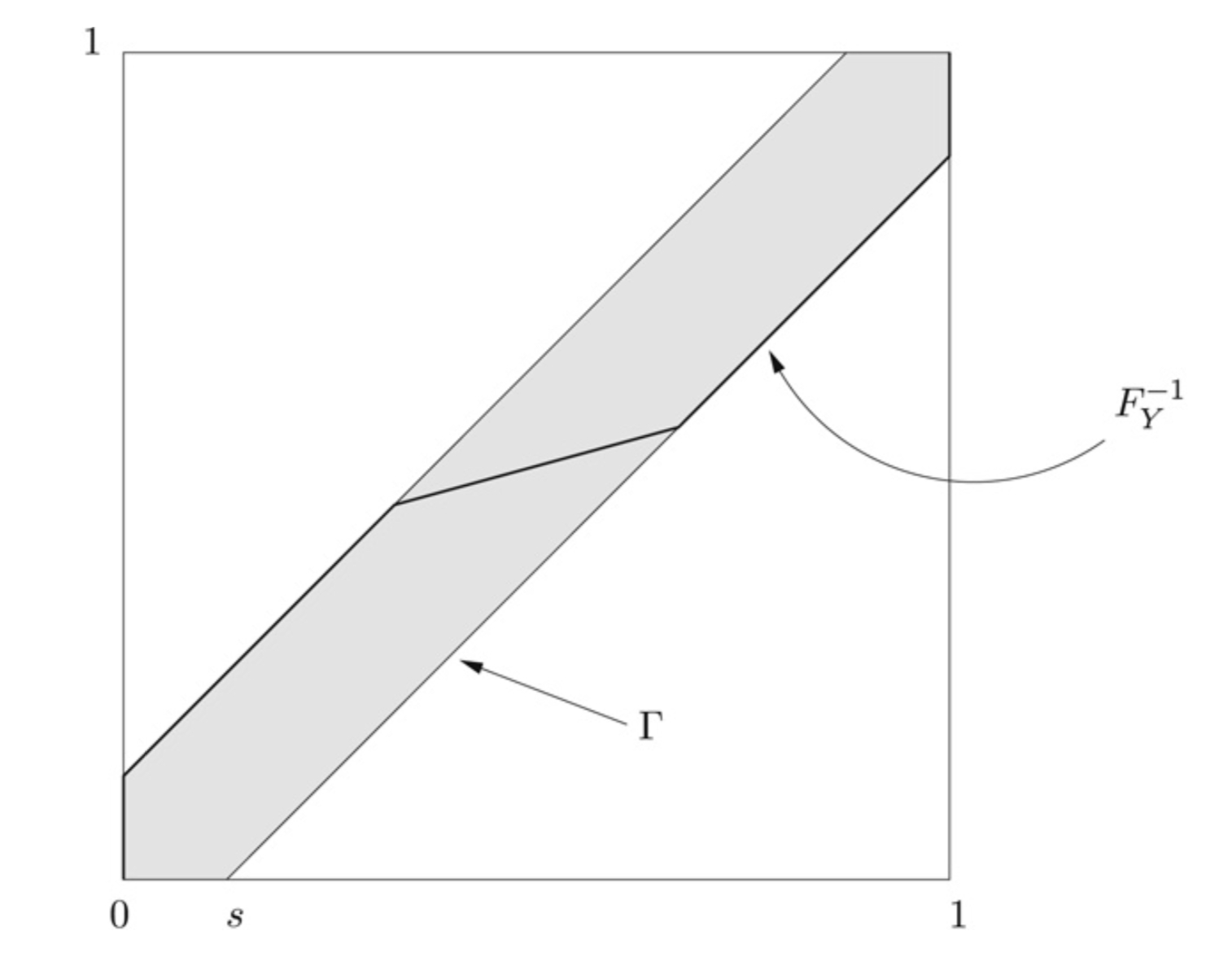}
\caption{The correspondence $\Gamma$ is given by the shaded area,
and the thick lines trace the inverse cumulative distribution
function of $Y$.} \label{MonteCarlo}
\end{center}
\end{figure}
\vskip6pt

We perform 1000 repetitions of the following testing procedure, and
we report the proportions of rejections out of
these 1000 repetitions. We first generate\footnote{We use MATLAB version 7.1 with random seed 777.} a sample
$(U_1,\ldots,U_n)$ of iid uniform $[0,1]$, with $n=100,500,1000$ and
compute the sample of observable variables $(Y_1,\ldots,Y_n)$ as
$(F^{-1}(U_1),\ldots,F^{-1}(U_n))$. $P_n$ is the empirical law of
$(Y_1,\ldots,Y_n)$, and $\mathcal{C}_{n,h_n}$ is the collection of
sets of the form $[0,Y_i]$, $i=1,\ldots,n$ with
$P_n[0,Y_i]=(1/n)\sum_{j=1}^n1_{\{Y_j\leq Y_i\}} \geq
\nu(\Gamma([0,Y_i]))-h_n=\min[1,Y_i+s]-h_n$ or
$[Y_i,1]$, $i,\ldots,n$ with $P_n[Y_i,1] \geq
\nu(\Gamma([Y_i,1]))-h_n=\min[1,1-Y_i+s]-h_n$.

For each sample, we draw 1000 bootstrap samples
$(Y_1^b,\ldots,Y_n^b)$, and call $P^b$ the law of the bootstrap
sample. For each bootstrap sample, we calculate the maximum of the
quantities $P^b[0,Y_i]-P_n[0,Y_i]$ for all $i$ such that
$[0,Y_i]\in\mathcal{C}_{n,h_n}$ and $P^b[Y_i,1]-P_n[Y_i,1]$ for all
$i$ such that $[Y_i,1]\in\mathcal{C}_{n,h_n}$, and call this maximum
$\max\mathbb{G}^b$. Order the $\max\mathbb{G}^b$ obtained for all
bootstrap draws, and call $c_\ast^\alpha$ the $(1-\alpha)1000$
largest, for $\alpha=0.01,0.05,0.1$. Reject if $c_\ast^\alpha$ is
smaller than the maximum of the quantities $P_n[0,Y_i]$ and
$P_n[Y_i,1]$ for $i=1\ldots,n$.

\begin{table}
\vskip15pt \caption{Rejection levels for the partially identified
case.}
\label{table: partially identified}\begin{center}
\begin{tabular}{l||c|c|c}
{Sample Size} &100&500&1000\\ \hline\hline {$\alpha=0.01$}
&0.001&0.007&0.008 \\ \hline {$\alpha=0.05$} &0.010&0.024&0.029 \\
\hline {$\alpha=0.10$} &0.029&0.049&0.066
\end{tabular}
\end{center}
\end{table}

\begin{table}
\vskip15pt \caption{Rejection levels for the exactly identified
case} \label{table: identified}\begin{center}
\begin{tabular}{l||c|c|c}
{Sample Size} &100&500&1000\\ \hline\hline {$\alpha=0.01$}
&0.019&0.024&0.014 \\ \hline {$\alpha=0.05$} &0.074&0.079&0.050 \\
\hline {$\alpha=0.10$} &0.138&0.135&0.105
\end{tabular}
\end{center}
\end{table}

\begin{table}
\vskip15pt \caption{Sensitivity of rejection levels to the
choice of tuning parameters} \label{table: tuning sensitivity}\begin{center}
\begin{tabular}{l||c|c|c|c|c|c}
{Sample Size} &\multicolumn{2}{c|}{100}&\multicolumn{2}{c|}{500}&\multicolumn{2}{c}{1000}\\
Tuning&$h_n=0.05$&$h_n=0.15$&$h_n=0.02$&$h_n=0.10$&$h_n=0.01$&$h_n=0.07$\\
\hline\hline {$\alpha=0.01$}
&0.004&0&0.012&0.002&0.019&0.005 \\ \hline {$\alpha=0.05$} &0.026&0.006&0.049&0.017&0.058&0.022 \\
\hline {$\alpha=0.10$} &0.064&0.020&0.090&0.034&0.111&0.043
\end{tabular}
\end{center}
\end{table}

\begin{figure}[h]
\begin{center}
\includegraphics[width=8cm]{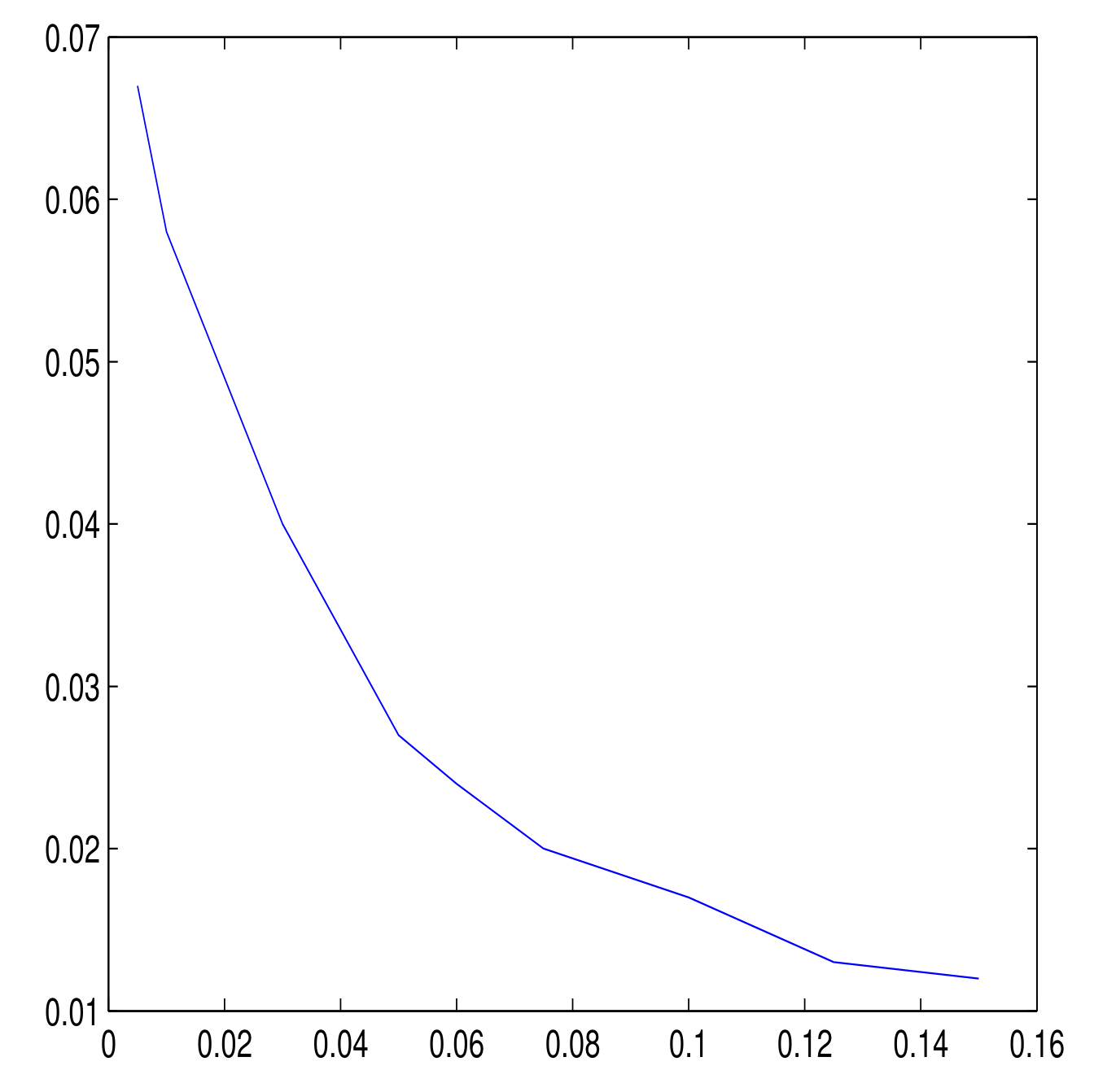}
\caption{\label{figure: tuning} Sensitivity to the tuning parameter.
Sample size 500, level 0.05, tuning parameter ranging from 0.005 to
0.15 on the $X$ axis, and rejections rates on the $Y$ axis.}
\end{center}
\end{figure}

The results are given in table~4 for the partially identified case
($s=0.15$) and in table~5, we give the benchmark of the exactly
identified case ($s=0$ and $h_n=1$), so that the test is a
traditional Kolmogorov-Smirnov specification test. The results are
given for $h_n$ on the boundary of the admissible rate, i.e.
$h_n=\sqrt{\ln\ln n/n}$. This rate was chosen as a power maximizing
rate (the rate that will ensure smaller quantiles, hence larger
rejection rates). This is the only justification for a choice of
rate that we can provide at this stage, as optimal rate choice is
beyond the scope of this paper. In applications, it is recommended
to provide results for different choices of rates, as one would
typically do in density, nonparametric regression or spectral
estimation. The rejection rates are low for small sample sizes and
improve sharply when sample size increases. To give a sense of the
sensitivity of rejection rates to the choice of the tuning parameter
$h_n$, table~6 reports rejection rates in the case of
$\alpha=0.01,0.05,0.1$ and $n=100,500,1000$ and choices of tuning
parameter $h_n$ that are significantly above, and significantly
below the initial choice of $h_n=\sqrt{\ln\ln n/n}$. For $n=1000$,
$\sqrt{\ln\ln n/n}=0.044$, so we report results for
$h_n=0.010,0.070$. For $n=500$, $\sqrt{\ln\ln n/n}=0.060$, so we
report results for $h_n=0.020,0.100$. For $n=100$, $\sqrt{\ln\ln
n/n}=0.120$, so we report results for $h_n=0.050,0.150$. Notice that
we decrease the investigated range of tuning parameter with sample
size, which corresponds to the fact that the tuning parameter
converges to zero. For $n=100$, the rejection rates are sensitive to
the choice of rate within the theoretical range
(assumption~\ref{assumption: bandwidth}) of tuning parameters. For
$n=500$, there is still sensitivity to the choice of $h_n$, somewhat
less so for $n=1000$. However, as in the case of bandwidth in kernel
estimation or in local spectral estimation of time series, it is
highly recommended to report empirical results with a good range of
values of the tuning parameter $h_n$. Figure~\ref{figure: tuning}
graphs rejections rates against tuning parameter to give a better
sense of this sensitivity for sample size 500 and level 0.05. It is
important also to note that higher values of the tuning parameter
lead to less filtering, i.e. more sets are used in the computation
of the supremum of the bootstrap empirical process, leading to
larger quantiles, hence smaller rejection rates. Hence it also shows
how crucial the filtering procedure is, since without it, the power
of the test would be very poor.

\section*{Conclusion}

We propose a test of the specification of a structure in the sense
of \cite{KR:50}, extended by \cite{Jovanovic:89}, where
observable variables and latent variables are related by a
many-to-many mapping, thereby allowing censored observable variables
and multiple equilibria. We apply mass transportation duality to
derive a simple necessary and sufficient condition for compatibility
of such structures and data in complete generality, and to justify
the use of a generalized Kolmogorov-Smirnov test statistic. We
propose a generically applicable and easily implementable procedure
to test compatibility of structure and data, and to construct
confidence regions for partially identified parameters specifying
the structure. This work therefore complements other proposals,
which tend to focus on models defined by moments inequalities. The
small sample performance of the test is investigated in a Monte
Carlo experiment, and is found to be comparable to the performance
of the traditional Kolmogorov-Smirnov specification test statistic.

\section*{Appendix}

\subsection*{Additional definitions}
\begin{definition} \label{definition: measurable correspondence}
A many-to-many mapping $\Gamma:$
$\mathbb{R}^{d_1}\rightrightarrows\mathbb{R}^{d_2}$ is called
measurable if for each open set ${\cal O}\subseteq\mathbb{R}^{d_2}$,
$\;\Gamma^{-1}({\cal O})=\{x\in\mathbb{R}^{d_1}\;|\;
\Gamma(x)\cap{\cal O}\neq\varnothing\}$ is a measurable subset of
$\mathbb{R}^{d_1}$.
\end{definition}

\begin{definition} \label{definition: Hausdorff metric}
Calling $d$ the Euclidean metric, the Hausdorff metric $d_H$ between
two sets $A_1$ and $A_2$ is defined by
\begin{eqnarray*}d_H(A_1,A_2)=\max\left( \sup_{y\in A_1}\inf_{z\in
A_2}d(y,z),\sup_{z\in A_2}\inf_{y\in
A_1}d(y,z)\right).\end{eqnarray*}
\end{definition}

\subsection*{Proofs of results in the main text}

\begin{varproof}[Proof of proposition \protect\ref{proposition: Monge-Kantorovich}]:
Since $\Gamma$ is closed valued,
$\varphi(y,u)=1_{\{u\notin\Gamma(y)\}}$ is lower semicontinuous, so
that we can apply lemma~\ref{lemma: Monge-Kantorovich} below to
yield \begin{eqnarray}\inf_{\pi\in{\cal M}(P,\nu)}\;\pi\varphi
=\sup_{f\oplus g\leq\varphi}\;(Pf+\nu g),\label{dual
Monge-Kantorovich with functions}\end{eqnarray} where $f\oplus
g\leq\varphi$ stands for $f(y)+g(u)\leq\varphi(y,u)$ all $y,u$.
Since the sup-norm of the cost function is 1 (the cost function is
an indicator), the supremum in (\ref{dual Monge-Kantorovich with
functions}) is attained by pairs of functions $(f,g)$ in ${\cal F}$,
defined by
\begin{eqnarray*}{\cal
F}=\{(f,g)\in\mathbb{L}^1(P)\times\mathbb{L}^1(\nu),\;0\leq
f\leq1,\;-1\leq
g\leq0,\\f(y)+g(u)\leq1_{\{u\notin\Gamma(y)\}},\;\mbox{$f$ upper
semicontinuous}\}.\end{eqnarray*} Now, $(f,g)$ can be written as a
convex combination of pairs $(1_A,-1_B)$ in ${\cal F}$. Indeed,
$f=\int_{0}^{1}1_{\{f\geq x\}}\,dx$ and $g=\int_{0}^{1}-1_{\{g\leq
-x\}}\,dx$, and for all $x$, $1_{\{f\geq
x\}}(y)-1_{\{g\leq-x\}}(u)\leq1_{\{u\notin\Gamma(y)\}}$. Since the
functional on the right-hand side of (\ref{dual Monge-Kantorovich
with functions}) is linear, the supremum is attained on such a pair
$(1_A,-1_B)$. Hence, the right-hand side of (\ref{dual
Monge-Kantorovich with functions}) specializes to
\begin{eqnarray}\sup_{A\times B\subseteq D}
(P(A)-1+\nu(B)).\label{duality with indicators}\end{eqnarray} For
$D=\{(y,u):\, u\notin\Gamma(y)\}$, $A\times B\subseteq D$ means
that if $y\in A$ and $u\in B$, then $u\notin\Gamma(y)$. In other
words $u\in B$ implies $u\notin\Gamma(A)$, which can be written
$B\subseteq\Gamma(A)^{c}$. Hence, the dual problem can be written
\begin{eqnarray*}\sup_{\Gamma(A)\subseteq B^c}
(P(A)-1+\nu(B))=\sup_{\Gamma(A)\subseteq B}
(P(A)-\nu(B)).\end{eqnarray*} and the result follows immediately.
\end{varproof}

\begin{lemma}\label{lemma: Monge-Kantorovich}
If $\varphi:{\cal Y}\times{\cal U}\rightarrow\mathbb{R}$ is bounded,
non-negative and lower semicontinuous, then
\begin{eqnarray*}\inf_{\pi\in{\cal M}(P,\nu)}\;\pi\varphi
=\sup_{f\oplus g\leq\varphi}\;(Pf+\nu g).\end{eqnarray*}
\end{lemma}

\begin{varproof}[Proof of lemma \protect\ref{lemma: Monge-Kantorovich}]:
The left-hand side is immediately seen to be always larger than
the right-hand side, so we show the reverse inequality. It is a
specialization of the Monge-Kantorovich duality to zero-one cost, which can
also be proved using Proposition (3.3) page 424 of
\cite{Kellerer:84}, but we give a direct proof due
to N. Belili for completeness. \vskip6pt
[a] case where $\varphi$ is continuous and ${\cal U}$ and ${\cal
Y}$ are compact.\\Call $G$ the set of functions on ${\cal
Y}\times{\cal U}$ strictly dominated by $\varphi$ and call $H$ the
set of functions of the form $f+g$ with $f$ and $g$ continuous
functions on ${\cal Y}$ and ${\cal U}$ respectively. Call
$s(c)=Pf+\nu g$ for $c\in H$. It is a well defined linear
functional, and is not identically zero on $H$. $G$ is convex and
sup-norm open. Since $\varphi$ is continuous on the compact ${\cal
Y}\times{\cal U}$, we have
\begin{eqnarray*}s(c)\leq\sup f+\sup g<\sup\varphi\end{eqnarray*} for
all $c\in G\cap H$, which is non empty and convex. Hence, by the
Hahn-Banach theorem, there exists a linear functional $\eta$ that
extends $s$ on the space of continuous functions such that
\begin{eqnarray*}\sup_{G}\;\eta=\sup_{G\cap H}\;s.\end{eqnarray*}
By the Riesz representation theorem, there exists a unique finite
non-negative measure $\pi$ on ${\cal Y}\times{\cal U}$ such that
$\eta(c)=\pi c$ for all continuous $c$. Since $\eta=s$ on $H$, we
have \begin{eqnarray*}\int_{{\cal Y}\times{\cal U}}
f(y)\;d\pi(y,u)&=&\int_{\cal Y}f(y)\;dP(y)\\
\int_{{\cal Y}\times{\cal U}} g(u)\;d\pi(y,u)&=&\int_{\cal
Y}g(u)\;d\nu(y),\end{eqnarray*} so that $\pi\in{\cal M}(P,\nu)$ and
\begin{eqnarray*}\sup_{f\oplus g\leq\varphi} (Pf+\nu g)=\sup_{G\cap
H} s=\sup_{G} \eta=\pi\varphi.\end{eqnarray*} \vskip8pt [b] ${\cal
Y}$ and ${\cal U}$ are not necessarily compact, and $\varphi$ is
continuous.\vskip6pt For all $n>0$, there exists compact sets $K_n$
and $L_n$ such that
\begin{eqnarray*} \max\left(P({\cal Y}\backslash K_n),\nu({\cal
U}\backslash L_n)\right)\leq\frac{1}{n}.\end{eqnarray*} Let
$(a,b)$ be an element of ${\cal Y}\times{\cal U}$ and define two
probability measures $\mu_n$ and $\nu_n$ with compact support by
\begin{eqnarray*}\mu_n(A)&=&P(A\cap K_n)+P(A\backslash K_n)\delta_a(A)\\
\nu_n(B)&=&\nu(B\cap L_n)+\nu(B\backslash
L_n)\delta_b(B),\end{eqnarray*} where $\delta$ denotes the Dirac
measure. By [a] above, there exists $\pi_n$ with marginals $\mu_n$
and $\nu_n$ such that
\begin{eqnarray*}\pi_n\varphi\leq\sup_{f\oplus g\leq\varphi}
(Pf+\nu g)+\frac{\varphi(a,b)}{n}.\end{eqnarray*} Since $(\pi_n)$
has weakly converging marginals, it is weakly relatively compact.
Hence it contains a weakly converging subsequence with limit
$\pi\in{\cal M}(P,\nu)$. By Skorohod's almost sure representation
(see for instance theorem 11.7.2 page 415 of
\cite{Dudley:2002}), there exists a sequence of random
variables $X_n$ on a probability space $(\Omega,{\cal
A},\mathbb{P})$ with law $\pi_n$ and a random variable $X_0$ on
the same probability space with law $\pi$ such that $X_0$ is the
almost sure limit of $(X_n)$. By Fatou's lemma, we then have
\begin{eqnarray*}\mbox{liminf}\;\pi_n\varphi
= \mbox{liminf}\, \mathbb{E} \varphi(X_n) \geq \mathbb{E}
\,\mbox{liminf} \varphi(X_n) = \mathbb{E} \varphi(X_0) =
\pi\varphi.\end{eqnarray*} Hence we have the desired result.
\vskip8pt [c] General case.\vskip6pt $\varphi$ is the pointwise
supremum of a sequence of continuous bounded functions, so the
result follows from upward $\sigma$-continuity of both
$\inf_{\pi\in{\cal M}(P,\nu)}\pi\varphi$ and $\sup_{f\oplus
g\leq\varphi} (Pf+\nu g)$ on the space of lower semicontinuous
functions, shown in propositions (1.21) and (1.28) of
\cite{Kellerer:84}.
\end{varproof}

\begin{varproof}[Proof of theorem \protect\ref{theorem: size}]:
We show that $T_n$ and $\tilde{T}_n$ converge in distribution
(notation $\rightsquigarrow$) to the same limit, which has a
continuous distribution function. Hence, the result follows.
\begin{itemize} \item Case where $Y=D$ discrete.
Let $A_0$ be the subset of $\mathfrak{X}_D$ that achieves the
maximum of $\delta(A)=P(A)-\nu(\Gamma(A))$ over $A\in{\cal
C}\backslash{\cal C}_b$. Call $\delta_0=\delta(A_0)$, and note that
$\delta_0<0$. We have \begin{eqnarray*} T_n &=&
\sup_{A\in2^{\mathfrak{X}_D}} [\mathbb{G}_n(A) + \sqrt{n}(P(A)
-\nu(\Gamma(A)))]\\ &=& \max\{ \sup_{\mathcal{C}_b} \mathbb{G}_n ,
\sup_{A\in2^{\mathfrak{X}_D}\backslash\mathcal{C}_b}
[\mathbb{G}_n(A) + \sqrt{n}(P(A) - \nu(\Gamma(A)))] \}.
\end{eqnarray*} The second term in the maximum of the preceding
display is dominated by
\[\sup_{2^{\mathfrak{X}_D}\backslash\mathcal{C}_b} \mathbb{G}_n + \sqrt{n}
\delta_0,\]whose limsup is almost surely non-positive. Hence
$T_n\rightsquigarrow \sup_{\mathcal{C}_b}\mathbb{G}$ follows from
the convergence of the empirical process.
$\tilde{T}_n\rightsquigarrow \sup_{\mathcal{C}_b} \mathbb{G}$
follows from the fact that, under assumption~\ref{assumption:
bandwidth}, for all $n$ sufficiently large, ${\cal C}_{h_n}$ is
almost surely equal to ${\cal C}_b$.

\item Case of $Y=C$ absolutely continuous with respect to Lebesgue measure.
Consider two sequences of positive numbers $l_n$ and $h_n$ such that
they both satisfy assumption~\ref{assumption: bandwidth}, $l_n>h_n$
and $(l_n-h_n)^{-1}\sqrt{\frac{\ln\ln n}{n}}\rightarrow0$. Notice
that $\{\varnothing,{\mathbb{R}^{d_C}}\} \subseteq
\mathcal{C}_b,\mathcal{C}_{b,h},\mathcal{C}_{h}$ for any $h>0$.
Since $\mathbb{G}_n({\mathbb{R}^{d_C}})=0$, we therefore have
$\sup_{\mathcal{C}_b}\mathbb{G}_n$,
$\sup_{\mathcal{C}_{b,l_n}}\mathbb{G}_n$ and
$\sup_{\mathcal{C}_{h_n}}\mathbb{G}_n$ non-negative. Hence, calling
$\zeta_n$ the indicator function of the event
$\sup_\mathcal{C}\mathbb{G}_n\leq(l_n-h_n)\sqrt{n}$, we can write
\begin{eqnarray*}\zeta_n \sup_{\mathcal{C}_b}\mathbb{G}_n&\leq&
\zeta_n \max\left\{\sup_{\mathcal{C}_b}[\mathbb{G}_n+\sqrt{n}(P-\nu\Gamma)],
\sup_{\mathcal{C}\backslash\mathcal{C}_b}[\mathbb{G}_n+\sqrt{n}(P-\nu\Gamma)]\right\}\\
&\leq&\zeta_n T_n\\
&\leq&\zeta_n \sup_{\mathcal{C}_{h_n}}\mathbb{G}_n\\
&\leq&\zeta_n\sup_{\mathcal{C}_{b,l_n}}\mathbb{G}_n,\end{eqnarray*} where the first inequality
holds because the left-hand side is equal to the first term in the
right-hand side, the second inequality holds trivially as an
equality since $\mathcal{C}=\mathcal{C}_b\cup\mathcal{C}\backslash\mathcal{C}_b$,
the third inequality holds because on $\mathcal{C}\backslash\mathcal{C}_{h_n}$, we have by definition
$\mathbb{G}_n + \sqrt{n}(P-\nu\Gamma) = \sqrt{n}(P_n-\nu\Gamma)
\leq -h_n \leq 0$, and the last inequality holds because on
$\{\zeta_n=1\}$, we have that $A\in\mathcal{C}_{h_n}$ implies
$\nu\Gamma(A) \leq P_n(A) + h_n = P(A) + (P_n-P)(A) +h_n \leq P(A)
+ \sup_\mathcal{C}\mathbb{G}_n/\sqrt{n} + h_n \leq P(A) + l_n - h_n +
h_n = P(A) + l_n $, which implies that $A\in\mathcal{C}_{b,l_n}$.

By lemma~\ref{lemma: Einmahl and Mason} and Theorem 2.5.2 page 127
of \cite{VW:96}, we have that both
$\sup_{\mathcal{C}_b}\mathbb{G}_n$ and
$\sup_{\mathcal{C}_{b,l_n}}\mathbb{G}_n$ converge in distribution to
$\sup_{\mathcal{C}_b}\mathbb{G}$. It is shown below that
$\zeta_n\rightarrow_p1$, so that Slutsky's lemma (lemma 2.8 page 11
of \cite{Vaart:98}) yields the weak convergence of
$\zeta_n\sup_{\mathcal{C}_b}\mathbb{G}_n$ and
$\zeta_n\sup_{\mathcal{C}_{b,l_n}}\mathbb{G}_n$ to the same limit,
and hence that of $\zeta_n T_n$ and
$\zeta_n\sup_{\hat\mathcal{C}_{h_n}}\mathbb{G}_n$. It follows from
Slutsky's lemma again that
\begin{eqnarray*}T_n\rightsquigarrow
\sup_{\mathcal{C}_b}
\mathbb{G}\hskip10pt\mbox{and}\hskip10pt\tilde{T}_n\rightsquigarrow \sup_{\mathcal{C}_b}
\mathbb{G}.\end{eqnarray*}

We now prove that
$\zeta_n\rightarrow_p1$. Indeed, for any $\epsilon>0$, $P(
|\zeta_n-1| > \epsilon ) = P( \zeta_n = 0 ) = P( \sup_\mathcal{C}\mathbb{G}_n > (l_n-h_n)\sqrt{n} )
\rightarrow 0$ by the Law of the Iterated Logarithm (see 12.5 page 476 of \cite{Dudley:2002}),
since $(l_n-h_n)\sqrt{n}\gg\sqrt{\ln\ln n}$ by assumption.

\end{itemize}
\end{varproof}

\begin{lemma} We have
\begin{eqnarray*}\sup_{A\in{\cal C}_{b,h_n}}\mathbb{G}_n(A)
\rightsquigarrow\sup_{A\in{\cal C}_b}\mathbb{G}(A),\end{eqnarray*}
\label{lemma: Einmahl and Mason}\end{lemma}

\begin{varproof}[Proof of lemma \protect\ref{lemma: Einmahl and Mason}]:
Take a bandwidth sequence $l_n$ that satisfies
assumption~\ref{assumption: bandwidth}, and take
$\mathcal{C}_{b,l_n}$ as in table~3. Under
assumption~\ref{assumption: frank separation}, take $A\in
\mathcal{C}_{b,l_n}$ and an $A_b\in\mathcal{C}_b$ such that
$d_H\left( A,A_b\right) \leq \zeta_n=Kl_{n}^{\eta}$ (we suppress the
dependence of $A_b$ on $A$ for ease of notation). As
$\mathcal{C}_{b}\subseteq\mathcal{C}_{b,l_n}$, one has
\begin{equation}
\sup_{A\in \mathcal{C}_b}\mathbb{G}_n(A)\leq \sup_{B\in\mathcal{C}_{b,l_n}}\mathbb{G}_{n}(A)  \label{one}
\end{equation}
Second, since $A_b\subseteq A$, one has
\begin{eqnarray*}\sup_{A\in
\mathcal{C}_{b,l_{n}}}\mathbb{G}_{n}(A)&=&\sup_{A\in \mathcal{C}
_{b,l_{n}}}\left[ \mathbb{G}_{n} (A_b) +
\mathbb{G}_{n}(A\backslash A_b) \right]\\&\leq& \sup_{A\in
\mathcal{C} _{b,l_{n}}}\left[ \mathbb{G}_{n} (A_b)\right] +
\sup_{A\in \mathcal{C} _{b,l_{n}}}\left[\mathbb{G}_{n}(A\backslash
A_b) \right].\end{eqnarray*}  If we have that
\begin{eqnarray*}
\sup_{A\in\mathcal{C}_{b,l_n}}\left\vert \mathbb{G}_{n}(A\backslash
A_b) \right\vert = O_{\mathrm{a.s.}}\left( \sqrt{ \zeta_{n}\ln\ln
n }\right),
\end{eqnarray*}
then
\begin{equation}
\sup_{A\in \mathcal{C}_{b,l_{n}}}\mathbb{G}_{n}(A)=\sup_{A\in
\mathcal{C}_{b,l_{n}}}\left[ \mathbb{G}_{n}(A_b) \right] +
O_{\mathrm{a.s.}}\left( \sqrt{\zeta_{n}\ln\ln n}\right)
\label{two}
\end{equation} noting the dependence of $A_b$ on $A$ in the
expression above. But since $A_b \in \mathcal{C}_{b}$, one has
$\sup_{A\in \mathcal{C}_{b,l_{n}}}\left[ \mathbb{G}_{n}\left(
A_b\right) \right] \leq \sup_{A\in
\mathcal{C}_{b}}\mathbb{G}_{n}(A)$. This fact, along with
(\ref{one}) and (\ref{two}), yields the result.\vskip6pt

We now show that we have indeed that \begin{eqnarray*}
\sup_{A\in\mathcal{C}_{b,l_n}}\left\vert \mathbb{G}_{n}(A\backslash
A_b) \right\vert = O_{\mathrm{a.s.}}\left( \sqrt{ \zeta_{n}\ln\ln
n }\right). \end{eqnarray*}

This relies on the construction of a local empirical process
relative to the thin regions $A\backslash A_b$. First consider
such a region. If $A\in\mathcal{C}_{b}$, the result holds trivially,
so that we may assume that $A\in\mathcal{C}_{b,l_n}\backslash\mathcal{C}_b$,
so that $A\backslash A_b$ is not empty. We distinguish the
case where $A$ is a bounded rectangle, and the cases where $A$ is
unbounded.\vskip6pt

\begin{itemize}

\item[(i)] $A$ is a bounded rectangle, i.e. of the form
$(y_1,z_1)$ $\times$ $\ldots$ $\times$ $(y_{d_y},z_{d_y})$, with
$y_1,$ $\ldots,$ $y_{d_y},z_1,$ $\ldots,$ $z_{d_y}$ real. Then,
since $d_{H}(A,A_b)\leq \zeta_n$, $A_b$ is also a bounded
rectangle, and the $A\backslash A_b$ is the union of at least one
(since $A$ and $A_b$ are distinct) and at most $f(d_y)$ (the
number of faces of a rectangle in $\mathbb{R}^{d_y}$) rectangles
with at least one dimension bounded by $\zeta_n$.

\item[(ii)] $A$ is an unbounded rectangle, i.e. of the same form
as above, except that some of the edges are $+\infty$ of
$-\infty$. Then $A_b$ is also an unbounded rectangle, and
$A\backslash A_b$ is also the union of a finite number of
rectangles with one dimension bounded by $\zeta_n$.

\end{itemize}
In both cases $(i)$, and $(ii)$, $A\backslash A_b$ is the union of
a finite number of rectangles with at least one dimension bounded
by $\zeta_n$. Hence if we control the supremum of the empirical
process on one of these thin rectangles, when $A$ ranges over
$\mathcal{C}_{b,l_n}$, we can control it on $A\backslash
A_b$.\vskip6pt

Hence, it suffices to prove that \begin{eqnarray*} \sup_{A\in\mathcal{C}_{b,l_n}}
\left\vert \mathbb{G}_{n}(\varphi_n (A)) \right\vert =
O_{\mathrm{a.s.}}\left( \sqrt{ \zeta_n\ln\ln n }\right),
\end{eqnarray*} where $\varphi_n$ is the homothety that carries
$A$ into one of the thin rectangles described above.\vskip6pt

As an homothety, $\varphi_n$ is invertible and bi-measurable, and
since $\varphi_n(A)$ has at least one dimension bounded by
$\zeta_n$, and $P$ is absolutely continuous with respect to
Lebesgue measure, $P(\varphi_n(A))=O(\zeta_n)$ uniformely when $A$
ranges over $\mathcal{C}_{b,l_n}$. Now, for any $A\in\mathcal{C}_{b,l_n}$, we have
\begin{eqnarray*}\mathbb{G}_{n}(\varphi_n
(A))&=&\sqrt{n}\left[P_n(\varphi_n(A))-P(\varphi_n(A))\right]\\&=&\frac{1}{\sqrt{n}}
\sum_{i=1}^{n}\left(
1_{\{\varphi_n(A)\}}(Y_i)-\mathbb{E}_{P}(1_{\{\varphi_n(A)\}}(Y))\right)\\
&=& \frac{1}{\sqrt{n}} \sum_{i=1}^{n}\left(
1_{A}(\varphi_n^{-1}(Y_i))-\mathbb{E}_{P}(1_{A}(\varphi_n^{-1}(Y)))\right)\\
&:=&\sqrt{\zeta_n} L_n(1_A,\varphi_n),\end{eqnarray*}

where $L_n(1_A,\varphi_n)$ is defined as
\begin{eqnarray*}\frac{1}{\sqrt{n \zeta_n}} \sum_{i=1}^{n}\left(
1_{A}(\varphi_n^{-1}(Y_i))-\mathbb{E}_{P}(1_{A}(\varphi_n^{-1}(Y)))\right)\end{eqnarray*}
to conform with the notation of \cite{EM:97}.\vskip6pt

Conditions~A(i)-A(iv) of the latter hold for $a_n=b_n=l_n$ and $a=0$
under assumption~\ref{assumption: bandwidth}, and
conditions~S(i)-S(iii) and F(ii) and F(iv)-F(viii) hold because
${\cal F}$ is here the class of indicator functions of
$\mathcal{C}_{b,l_n}$, hence Donsker (see for instance example 2.6.1
page 135 of \cite{VW:96}). Hence Theorem~1.2 of
\cite{EM:97} holds, and
\begin{eqnarray*}\sup_{A\in\mathcal{C}_{b,l_n}}\left\vert L_n(1_A,\varphi_n)\right\vert
= O_{\mathrm{a.s.}}\left( \sqrt{ \ln\ln n }\right)\end{eqnarray*}
so that the desired result holds.

\end{varproof}

\begin{varproof}[Proof of theorem \protect\ref{theorem: feasible
procedure}]:
By theorem~2.4 page 857 of \cite{GZ:90}, the bootstrapped
empirical process $\mathbb{G}^\ast$ converges weakly to $\mathbb{G}$
conditionally almost surely, so that
\[\sup_{A\in\mathcal{C}_{h_n}}\mathbb{G}_n(A) \;\mbox{ and }\;
\sup_{A\in\mathcal{C}_{h_n}}\mathbb{G}^\ast(A)\] have the same
continuous limit. There remains to show that $T_n$ and $\hat{T}_n$
have the same limit, and that
$\sup_{A\in\mathcal{C}_{n,h_n}}\mathbb{G}^\ast(A)=
\sup_{A\in\mathcal{C}_{h_n}}\mathbb{G}^\ast(A)$ so that the result
follows. The latter derives from the fact that $\mathbb{G}^\ast$
takes at most $n$ different values over $\mathcal{C}_{h_n}$ which
are exhausted on $\mathcal{C}_{n,h_n}$. We now prove the former.
First, notice that $\mathcal{C}_n\subseteq\mathcal{C}$ implies
$\hat{T}_n\leq T_n$.

\begin{itemize} \item Case where $Y=D$ discrete. In that case, there
is $n_0$ such that for all $n\geq n_0$, $\mathcal{C}_n=\mathcal{C}$,
and the result trivially follows.
\item Case where $Y=C\in\mathbb{R}^{d_y}$ has a density with respect to Lebesgue
measure. By Theorem~9.14 page 291 of \cite{Villani:2003},
there is existence of a one-to-one bi-measurable (i.e. both itself
and its inverse are measurable) and Lipschitz (with constant 1)
function $\phi:[0,1]^{d_y}\rightarrow\mathbb{R}^{d_y}$ such that
$Y=\phi(V)$ and $V$ is distributed uniformly on $[0,1]^{d_y}$
($\phi$ is called a generalized quantile transformation).

Hence, for any set $A\in\mathcal{C}$, we can write \[P_n(A)=
\frac{1}{n}\sum_{i=1}^{n}1_{\{Y_i\in A\}}=
\frac{1}{n}\sum_{i=1}^{n}1_{\{\phi(U_i)\in A\}}=
\frac{1}{n}\sum_{i=1}^{n}1_{\{U_i\in \phi^{-1}(A)\}}=
\lambda_n(\phi^{-1}(A)),\] where $\lambda_n$ denotes the empirical
law associated with an iid sample of uniformly distributed variables
on $[0,1]^{d_y}$.

We have
$\hat{T}_n-T_n=\sup_{A\in\mathcal{C}_n}[P_n(A)-\nu(\Gamma(A)]-
\sup_{A\in\mathcal{C}}[P_n(A)-\nu(\Gamma(A)]$. We show that for all
$\epsilon>0$, there is an $n_0$ such that for all $n>n_0$,
\[\sup_{y\in\mathbb{R}^{d_y}}\inf_{j\in\{1,\ldots,n\}}
\left\{(P_n(-\infty,Y_j]-P_n(-\infty,y])+(\nu(\Gamma(-\infty,y]))-\nu(\Gamma(-\infty,Y_j]))
\right\}<\epsilon\] and we can proceed similarly for sets of the
form $(-\infty,y]^c$. The proof of the latter proceeds in three
steps: \begin{itemize} \item By the results stated in the two
paragraphs following equation (1) page 919 of
\cite{Talagrand:94}, we have for any $\eta>0$
\[\sup_{y\in[0,1]^{d_y}}\min_{j\in\{1,\ldots,n\}}\vert\vert
v-V_j\vert\vert=O_{\mathrm{a.s.}}\left(n^{\eta-1/\max(2,d_y)}\right).\]
Since $\phi$ is Lipschitz, the latter implies that
\[\sup_{y\in\mathbb{R}^{d_y}}\min_{j\in\{1,\ldots,n\}}\vert\vert
y-Y_j\vert\vert=O_{\mathrm{a.s.}}\left(n^{\eta-1/\max(2,d_y)}\right).\]
\item Consider the mapping $y\mapsto j(y)$ which achieves the
minimum of $\vert\vert y-Y_{j(y)}\vert\vert$. B
assumption~\ref{assumption: continuity}, we have for $n$ large
enough,
$\sup_{y\in\mathbb{R}^{d_y}}(\nu(\Gamma((-\infty,Y_{j(y)}]))-\nu(\Gamma((-\infty,y])))<\epsilon/2$.
\item We have
$\sup_{y\in\mathbb{R}^{d_y}}(P(-\infty,y)-P(-\infty,Y_{j(y)}))<\epsilon/4$,
since the set $(-\infty,y)\backslash(-\infty,Y_{j(y)}]$ shrinks
uniformly.
\item By Theorem~2.3 page 367 of \cite{Stute:84}, we have $\sup_{A\subset\mathbb{R}^{d_y}}
(P_n(A)-P(A))<\epsilon/4$ for $n$ large enough, and the result
follows.
\end{itemize}\end{itemize}
\end{varproof}

\begin{varproof}[Proof of theorem \protect\ref{theorem:
consistency}]:
Under $\mathbb{H}_a$, there is a set $A_0$ in $\mathcal{C}$ such
that $P(A_0)>\nu(\Gamma(A_0))$. Now the test statistic is
\begin{eqnarray}T_n&=&\sqrt{n}\sup_{A\in\mathcal{C}}
[P_n(A)-\nu(\Gamma(A))]\nonumber\\&=&\sup_{A\in\mathcal{C}}[\mathbb{G}_n(A)+
\sqrt{n}(P(A)-\nu(\Gamma(A)))] \nonumber\\&\geq&
\mathbb{G}_n(A_0)+\sqrt{n}[P(A_0)-\nu(\Gamma(A_0))].\label{equation:
power}\end{eqnarray} Hence,
\begin{eqnarray*}T_n-\tilde{T}_n&=&\sqrt{n}\sup_{A\in\mathcal{C}}
[P_n(A)-\nu(\Gamma(A))]-\sup_{A\in\mathcal{C}_{h_n}}\mathbb{G}_n(A)\\
&\geq&\sqrt{n}\sup_{A\in\mathcal{C}}
[P_n(A)-\nu(\Gamma(A))]-\sup_{A\in\mathcal{C}}\mathbb{G}_n(A)\\
&\geq&\mathbb{G}_n(A_0)+\sqrt{n}[P(A_0)-\nu(\Gamma(A_0))]
-\sup_{A\in\mathcal{C}}\mathbb{G}_n(A),\end{eqnarray*} where the
first inequality follows from the fact that
$\mathcal{C}_{h_n}\subseteq\mathcal{C}$, and the second inequality
follows from~(\ref{equation: power}). Since
$P(A_0)>\nu(\Gamma(A_0))$, we have
$\sqrt{n}[P(A_0)-\nu(\Gamma(A_0))]\rightarrow\infty$. Hence, since
$\mathbb{G}_n(A_0)-\sup_{A\in\mathcal{C}}\mathbb{G}_n(A)$ is a tight
sequence (this can be derived for instance from exponential bounds in
2.14.9 and 2.14.10 page 246 of \cite{VW:96}), we have
$\mathbb{P}(T_n\geq c_n^\alpha)\rightarrow1$ for all $\alpha>0$.
\end{varproof}

\pagebreak
\markboth{References}{References}
\printbibliography

\end{document}